# The Low Frequency Sensitivity to Gravitational Waves for ASTROD

ANTONIO PULIDO PATON[1], WEI-TOU NI[1,2]

[1]*Purple Mountain Observatory, Chinese Academy of Sciences, Nanjing, 210008, China*

[2]*National Astronomical Observatories, Chinese Academy of Sciences, Beijing, 100012, China*

Email: antonio@pmo.ac.cn

ASTROD is a relativity mission concept encompassing multi-purposes. One of its main purposes is to detect gravitational waves sensitive to low-frequency band similar to LISA, but shifted to lower frequencies. In this aspect, ASTROD would complement LISA in probing the Universe and study strong-field black hole physics. Since ASTROD will be after LISA, in the Cosmic Vision time-frame 2015-2025, a ten-fold improvement over LISA's accelerometer noise goal would be possible, allowing us to test relativistic gravity to 1 ppb and improve the gravitational-wave sensitivity. In this paper, we address to this possible improvement, especially in the frequency range below 100 $\mu Hz$. We look into possible thermal noise improvement, magnetic noise improvement, spurious discharging noise improvement and local gravitational noise improvement. We discuss various possibilities of lower-frequency gravitational-wave responses and their significance to potential astrophysical sources.

# 1 Introduction

The existence of gravitational waves is one of the cornerstone predictions of General Theory of Relativity. The detection of gravitational waves presents a big experimental challenge to the gravitational community. Due to their weakly interacting nature, the observation of gravitational waves opens the possibility of understanding new astrophysical phenomena at very early epochs of our Universe. ASTROD (Astrodynamical Space Test of Relativity using Optical Devices) [1-6] is a relativity mission concept encompassing multi-purposes. The main purposes are to map the solar-system gravity, to measure the related solar-system parameters, to test relativistic gravity and to detect gravitational waves. For gravitational-wave detection, ASTROD will be sensitive to low-frequency band, similar to LISA [7], but shifted to lower frequency. Since ASTROD will be after

LISA, in the Cosmic Vision time-frame 2015-2025, a ten-fold improvement over LISA's accelerometer noise goal would allow us to test relativistic gravity to 1 ppb and improve the gravitational-wave detection sensitivity. Improving sensitivity at low frequency, below 100 μHz, has been discussed to be of great importance in the study of astrophysical phenomena like coalescence of black hole binaries at high red shift.

In the present paper we analyze ASTROD low frequency sensitivity and its impact in the study of these astrophysical sources. In section 2 we discuss briefly the disturbance reduction control loop model for the drag-free control system. In section 3 we review the acceleration disturbances affecting the spacecraft motion. In section 4 we discuss direct proof mass acceleration disturbances which are of diverse origins: magnetic, thermal, charging and sensor back action. In section 5 we briefly discuss spurious interactions leading to coupling of the spacecraft and the proof mass via stiffness. In section 6 we discuss the parameter requirements to achieve ASTROD noise target at 100 μHz. Finally in section 7 we discuss ASTROD spectral acceleration disturbance below 100 μHz, ASTROD gravitational wave response and its impact in the study of MBH binaries.

# 2 Gravitational Reference Sensor (GRS) control loop

To measure relativistic effects, including gravitational waves, the proof masses, acting as the interferometer mirrors, must follow an as pure as possible geodesic motion. To accomplish it we employ drag-free control. In the case of ASTROD a spacecraft structure shields two proof masses from external disturbances. To minimize acceleration disturbances both proof masses must be centered in their respective housings. A heuristic control loop model simplified to a spacecraft and a single proof mass can be used to infer how the disturbances affect the proof mass geodesic motion. The total proof mass acceleration disturbance will be the sum of direct disturbances acting on the proof mass and external disturbances coupled via PM-SC (proof mass-spacecraft) stiffness.

The proof mass acceleration disturbance can then be written as [8, 9, 10, 11],

$$f_p \approx X_{nr}(-K) + f_{np} + \left( f_{ns} + TN_t \right) K \frac{1}{\omega^2 u} \qquad (2.1)$$

where $X_{nr}$, $f_{np}$, $f_{ns}$, $K$ and $u$ are the readout sensitivity associated to the displacement sensor, the proof mass direct acceleration disturbances, the external disturbances (included thruster noise), the PM-SC stiffness and the SC control loop gain, respectively. External disturbances and sensor readout noise will contribute to the total proof mass acceleration noise via stiffness coupling. In the following sections we review the main disturbances affecting the proof mass with emphasis in the low frequency regime.

# 3 Spacecraft acceleration disturbances

The spacecraft shields the proof mass from external disturbances. Nevertheless these disturbances contribute to the total proof mass acceleration noise by stiffness coupling. Micrometeorite impacts, solar wind, solar radiation pressure and thruster noise, are some of the known disturbances affecting the spacecraft. The dominant external disturbance is due to solar radiation pressure. Above 100 μHz solar irradiance fluctuations follows the power law [12],

$$\frac{\delta W_0}{W_0} \approx 1.3 \times 10^{-3} \left( \frac{1\,\text{mHz}}{f} \right)^{1/3} \qquad (3.1)$$

Below that frequency solar irradiance fluctuations are bigger than the values obtained by using formula (3.1). This is expected because of the irradiance variability when approaching the solar rotation period of about 25 days. Solar irradiance fluctuations at frequencies $3 \times 10^{-5}$, $10^{-5}$ and $3 \times 10^{-6}$ Hz are $0.54 \times 10^{-2}$, $10^{-2}$ and $1.9 \times 10^{-2}$ respectively [13]. Assuming an effective SC area to be $A_{SC} \approx 5$ m$^2$, a SC mass of $M_{SC} \approx 350$ kg, and a total solar irradiance $W_0 \sim 5500$ W m$^{-2}$ we obtain that (see table 1),

$$f_{ns} \approx 6.8 \times 10^{-10} \left( \frac{1\,\text{mHz}}{f} \right)^{1/3} \qquad (3.2)$$

in units m s$^{-2}$ Hz$^{-1/2}$, for frequencies above 0.1 mHz. At frequencies $3 \times 10^{-5}$, $10^{-5}$ and $3 \times 10^{-6}$ Hz, the values of solar radiation pressure disturbances are $2.8 \times 10^{-9}$, $5.2 \times 10^{-9}$ and $9.9 \times 10^{-9}$ m s$^{-2}$ Hz$^{-1/2}$ respectively. We also have to consider thruster noise. In the case of LISA, thruster noise is aimed to be about a few μN Hz$^{-1/2}$, giving an acceleration disturbance of $\sim 2 \times 10^{-9}$ m s$^{-2}$ Hz$^{-1/2}$.

# 4 Proof mass direct acceleration disturbances

Direct proof mass acceleration disturbances are divided in two categories: a) environmental disturbances and b) sensor back action disturbances. Within the first category are direct impacts onto the proof mass, and disturbances of diverse origin: magnetic, charging and thermal. To monitor the relative displacement between the proof mass and the spacecraft, capacitive or optical sensing has to be employed. These sensors are also utilized to compensate for stray imbalance forces within the spacecraft. While capacitive sensing offers the advantage of having been successfully tested in space, optical sensing offers an important improvement on displacement sensitivity and a much lower back action force compared to capacitive sensing.

## 4.1 Direct impacts

Both cosmic rays and residual gas molecules deposit momentum in the proof mass inducing acceleration disturbances. These disturbances, denoted as $f_{CR}$ and $f_{RG}$ respectively, are shown in table 2. In the case of cosmic rays it has been considered an impact rate, λ, of 30 $s^{-1}$ for protons at incident energy $E_d$ = 200 MeV, giving an acceleration disturbance of $1.5 \times 10^{-18}$ m $s^{-2}$ $Hz^{-1/2}$. Residual gas molecules induce disturbances proportional to the square root of the residual pressure. This acceleration disturbance is of the order of $1.6 \times 10^{-16}$ m $s^{-2}$ $Hz^{-1/2}$, for a housing pressure of $10^{-6}$ Pa. That pressure value is the requirement for ASTROD, which is a factor 3 improvement with respect to the value assumed for LISA.

## 4.2 Magnetic disturbances

Magnetic disturbances arise due to the interaction of the test mass magnetic properties, i.e., remnant magnetic moment and susceptibility, with local sources of magnetic fields and the interplanetary magnetic field. Expressions of magnetic acceleration disturbances, $f_{m1}$, $f_{m2}$ and $f_{m3}$, are given in table 2. Both $f_{m1}$ and $f_{m3}$ are DC terms, while $f_{m2}$, proportional to the magnetic susceptibility, scales with frequency as $f^{-2/3}$, being one of the important sources of noise at low frequency. Using the values of proof mass magnetic properties and magnetic fields listed in table 5, these magnetic disturbances at 0.1 mHz are approx. $f_{m1} \approx 7.2 \times 10^{-18}$ m $s^{-2}$ $Hz^{-1/2}$, $f_{m2} \approx 2.0 \times 10^{-17}$ m s$^{-2}$ Hz$^{-1/2}$ and $f_{m3} \approx 3.2 \times 10^{-17}$ m s$^{-2}$ Hz$^{-1/2}$. A magnetic shielding factor $\xi_m \approx 10$ has been considered.

## 4.3 Thermal fluctuating acceleration disturbances

Temperature difference across the proof mass housing leads to differential pressure causing acceleration disturbances (radiometer and out gassing effect, thermal radiation pressure, etc). Thermal fluctuations will also cause deformations in the spacecraft structure inducing gravity gradients. Proof mass housing temperature fluctuations stem from solar irradiance and power dissipation in the amplifiers. Expressions of thermal disturbances and their values at different frequencies are summarized in table 2 and 3. Temperature variation across the proof mass housing is parameterized in terms of the optical bench temperature fluctuations, $\delta T_{OB}$, and a thermal shielding factor, $\xi_{TS}$. Above 0.1 mHz optical bench thermal fluctuations are mainly due to power dissipation of amplifiers which has been estimated to be about $3.0 \times 10^{-5}(1 \text{ mHz}/f)^{1/2}$ K Hz$^{-1/2}$ for the LISA configuration. Below 0.1 mHz solar irradiance is the dominant contribution [13]. All these disturbances increase at low frequencies.

Improvements in active/passive thermal isolation would allow us to further reduce disturbances due to radiometer, out gassing, and thermal radiation. A value for the optical bench thermal isolation factor, $\xi_{TS}$, of 150 has been considered. In the case of LISA $\xi_{TS}$ is considered to be in between 30-100 in the error estimates. Both radiometer and out gassing effect depend linearly on pressure. These acceleration noises can be suppressed by having better vacuum, leaving thermal radiation as the main thermal disturbance at low frequency, along with thermal induced gravity gradients.

## 4.4 Lorentz and sensor back action disturbances. Capacitive versus optical sensing

To monitor relative displacements between the proof mass and the spacecraft we can employ either capacitive or optical sensing. In what follows the analysis of disturbances associated to charge accrued by the proof mass and disturbances associated to sensor back action are discussed in detail for both cases.

### 4.4.1 Lorentz disturbances

Charge is accrued by the proof mass due to bombardment by both cosmic rays and solar particles. Because of moving along the interplanetary magnetic field the proof mass experiences a Lorentz force proportional to its charge and velocity. The Lorentz proof mass acceleration, to first order, is given by,

$$a_L = \frac{\bar{\dot{Q}}t}{m_P \xi_e}\vec{v} \times \vec{B}_{ip} + \frac{\bar{\dot{Q}}t}{m_P \xi_e}\vec{v} \times \delta\vec{B}_{ip} + \frac{\delta Q}{m_P \xi_e}\vec{v} \times \vec{B}_{ip} \qquad (3.3)$$

where we assume a linear charging process, i.e.,

$$Q = \bar{\dot{Q}}t + \delta Q(t) \qquad (3.4)$$

The first term in (3.3) gives rise to a coherent Fourier component while the second and third terms are acceleration disturbances associated to field and charge fluctuations，$f_{L1}$ and $f_{L2}$ in table 2, respectively.

### 4.4.2 Sensor back action disturbances. Capacitive sensing

Current missions under study like ASTROD I and LISA will employ capacitive sensing to measure the relative displacement between the proof mass and the spacecraft and also to exert forces and torques to actively control the proof mass-spacecraft relative position. In the case of employing capacitive sensing, back action disturbances stem from charge accrued by the proof mass and surrounding electrodes, and voltages applied for sensing and actuation. Other disturbances associated with capacitive sensing arise due to voltage quantization and dielectric losses. Capacitive sensing back action disturbances are listed in table 4.

Following [14], back action acceleration terms due to charge are given by,

$$a_{BA} \approx \frac{\bar{\dot{Q}}^2}{2C_T^2 m_P}\frac{\partial C_T}{\partial k}t^2 - \frac{\partial V_T}{\partial k}\frac{\bar{\dot{Q}}}{m_P}t + \frac{1}{m_P}\left[\frac{\bar{\dot{Q}}t}{C_T^2}\frac{\partial C_T}{\partial k} - \frac{\partial V_T}{\partial k}\right]\delta Q(t) \qquad (3.5)$$

where $C_T$ and $V_T$ are the capacitance and voltage of the proof mass. The first and second term in (3.5) give rise to coherent Fourier components, which will be discussed later on, while the third and fourth terms, are associated to charge fluctuation, $f_{\delta q,1}$ and $f_{\delta q,2}$ in table 4 respectively. Both charging disturbances, $f_{\delta q,1}$ and $f_{\delta q,2}$, decrease by increasing the gap. Disturbance $f_{\delta q,2}$ arise due to gap asymmetries across the proof mass because of machining inaccuracies. On the other hand the disturbance $f_{\delta q,1}$, could be suppressed by decreasing the potential difference across the proof mass, $V_d$, which for LISA noise estimates is considered to be $5 \times 10^{-3}$ V. Both $f_{\delta q,1}$ and $f_{\delta q,2}$ increase at low frequency as $1/f$.

Other back action disturbances arise due to fluctuation in voltage differences across opposite electrodes, $\delta V_d$ ($f_{\delta V,1}$ and $f_{\delta V,2}$). Dielectric losses in the surrounding electrodes also induces acceleration disturbances, depending on the loss angle, $\delta$, and average voltage applied to the electrodes. This disturbance also increases at low frequencies.

Another disturbance associated to capacitive sensing arises from the approximation to the quantization process. The net force exerted on the proof mass is given by [11],

$$F_{x0} \approx \frac{C_x}{2d}\frac{C_g}{C}\left(V_{x1}^2 - V_{x2}^2\right) - \frac{C_x}{d}\frac{q}{C}\left(V_{x1} - V_{x2}\right) \approx 2.5\times10^{-14}\left(\frac{V_d}{5\times10^{-3}}\right)\left(\frac{V_{x0}}{10^{-2}}\right)\text{N} \quad (3.6)$$

where we assume a discharged proof mass.

## 4.4.3 Coherent charging signals (CHS)

Time dependent signals due to charging (CHS), via Coulomb and Lorentz forces, give rise to coherent Fourier components. By inspection of (3.3) and (3.5), time dependent acceleration terms can be written as,

$$a_{CHS} = h_k(t) = (\Phi_k + \Theta_k)t + \Xi_k t^2 \quad (3.7)$$

where, if we consider a simplified model of parallel plate capacitors, we have that,

$$\Phi_k \approx \frac{\bar{\dot{Q}} v B_{ip}}{m_P \xi_e}; \quad \Theta_k \approx -\frac{\bar{\dot{Q}} C_x}{m_P C d} V_d; \quad \Xi_k \approx \frac{2C_x}{m_P}\left(\frac{\bar{\dot{Q}}}{Cd}\right)^2 \Delta d \quad (3.8)$$

The one-sided power spectral density of the signal (3.7) is defined by,

$$A_{CHS}^2 = \frac{2\left|H_k(f)\right|^2}{\tau} \quad (3.9)$$

where $H_k(f)$ is the Fourier transform of $h_k(t)$ and $\tau$ is the integration time. Expressions (3.8) show that coherent Fourier signals appear due to capacitance gap and voltage asymmetries across the proof mass, in the case of Coulomb forces, and the interplanetary magnetic field, in the case of the Lorentz force. Gap asymmetries depend on machining accuracy, while voltage offsets relate to the work function stability of the conductor domains.

In [14] it is shown that for typical parameter values these coherent charging signals can exceed the instrumental noise target for LISA. The signals associated to Coulomb forces are well over the acceleration noise target in the nominal LISA bandwidth while the signal associated to Lorentz force falls below the sensitivity goal for LISA assuming certain degree of electrostatic shielding. Moreover their signal peak's amplitudes increase with decreasing frequency.

Different approaches should be taken for the cases of using capacitive or optical sensing. In the case of using capacitive sensing all CHS have to be considered. Different ways of reducing these signals are discussed in [14]. Because these are proportional to the proof mass charging rate, an obvious way of reducing the magnitude of CHS would be to lower the net charging rate (via constant UV illumination). This itself will introduce new shot noise terms due to proof mass discharging, $f_{L2}^{disc}$ , $f_{\delta q,1}^{disc}$ and $f_{\delta q,2}^{disc}$ for which it has been assumed that $\delta q^{charging} \approx \delta q^{discharging}$.

If that is the case we will have a proof mass, up to a certain degree, free of charge and the acceleration disturbances associated to $f_{L1}$ , $f_{\delta q,2}$ and $f_{\delta V,2}$ , which are proportional to the total charge accrued, would be fully suppressed.

In the case of employing an optical sensor instead of capacitances, only the coherent signal associated to Lorentz force needs to be taken into account. Again the same approach as with capacitive sensing can be followed. Nevertheless because the CHS associated to Lorentz is much smaller than the ones associated to Coulomb forces, we can consider the proof mass to be periodically discharged. If the integration time, $\tau$, is much bigger than the discharging period, $T$, then the signals at $f \approx 0$ and $f \approx n/T$ are given by,

$$H_k^{ASTROD}(f) = \frac{T}{2}\Phi_k \tau \operatorname{sinc}(\tau f), \qquad f \approx 0 \tag{3.10}$$

$$H_k^{ASTROD}(f) = \frac{iT}{2\pi n}\Phi_k \tau \operatorname{sinc}\left\{\tau\left(f - \frac{n}{T}\right)\right\}, \qquad f \approx \frac{n}{T} \tag{3.11}$$

Using a effective charging rate of 288 $e^+$/s, $\xi_e = 100$, an orbital velocity of $4 \times 10^4$ m $s^{-1}$, an interplanetary magnetic field value of $1.2 \times 10^{-7}$ T and a proof mass of 1.75 kg, we obtain that $\Phi_k \approx 1.3\times10^{-21}$. Taking into account this value the acceleration noise sensitivity at 0.1 mHz is ~ $1.5 \times 10^{-12}$ m $s^{-2}$ $Hz^{-1/2}$, assuming discharging the proof mass once a day and integrating for one year. This will require electrostatic shielding factors $\xi_e \geq 5\times10^3$ to achieve acceleration disturbances below $3 \times 10^{-16}$ m s$^{-2}$ Hz$^{-1/2}$.

## 4.5 Optical sensing

Sensor readout sensitivity coupled via spacecraft-proof mass stiffness contributes to the acceleration noise spectrum. Inspection of (2.1) shows that the acceleration disturbance associated to readout noise is given by,

$$f_p \approx KX_{nr} \qquad (3.12)$$

where $K$ is the stiffness and $X_{nr}$ is the sensor readout sensitivity. According to (3.12) a drag-free system can be realized by either decoupling the proof mass as much as possible from the spacecraft using a low sensitivity controller or employing a high sensitivity controller with a high stiffness. In the case of capacitive sensing, the need for small gaps between metallic surfaces turn out into strong sensitivity to charge accrued, patch fields, magnetic impurities, surface imperfections, etc. One obvious solution to this problem could be the use of optical sensing. Optical sensing allows for large gaps and also presents advantages over capacitive sensing in terms of higher sensitivity and very low back action force. The sensitivity for an optical sensor shot-noise limited is given by,

$$X_{nr} \approx \frac{1}{\pi}\left(\frac{hc\lambda}{2\eta P}\right)^{1/2} \qquad (3.13)$$

According to (3.13) sub-picometer precision can in principle be achieved, using 1.5 μm laser wavelength, with such low powers as 1 μW. The back action force associated to this power is <2P/c ~ $6.7 \times 10^{-15}$ N. Considering a proof mass of 1.75 kg and a counter balance force of 1% we can get as low as $3.8 \times 10^{-17}$ m s$^{-2}$ in acceleration back action force. Efforts towards the realization of an optical GRS are being made by several research groups. In [15] an optical lever sensor is considered, sending a laser beam through a single-mode optical fiber to the surface of the test mass. Longitudinal displacements and rotations of the test mass are evaluated by measuring the displacement of the laser reflected on the position sensing device (PSD). The optical fiber is needed to reduce the beam jitter while, the ultimate source of noise would be shot noise. At the frequency of interest the main source of noise is due to current noise of the trans-impedance amplifier used to read the photo-detector signals. They achieve displacement sensitivities of the order of $10^{-9}$ m $Hz^{-1/2}$, which is the LISA requirement, above 5 mHz. In [16] is discussed a Gravitational Reference Sensor where the two cubical test masses of LISA are merged into a spherical test mass eliminating the inter-proof mass noise. In LISA due to the cubic proof mass shape, the proof mass orientation must be controlled by forcing, so in that sense LISA is a step further from an ideal drag-free GRS. They also consider all-reflective optical readout based on diffractive optics. By using all-reflective grating beam splitters to measure the internal gap distance, optical path errors due to refractive index changes induced by temperature (dn/dT) are eliminated. For many optical materials optical path errors due to refractive index change induced by temperature are ten times larger than those due to thermal expansion. Grating interferometry also has the advantage that one can separate the external from the internal interferometer used to measure the position of the spacecraft and proof mass. Therefore one can optimize both interferometers independently using different powers and wavelengths. Optical fibers are employed to deliver and receive optical radiation to and from the critical interferometer paths. That eliminates the need for an optical bench structure and detector and sources can be located remotely from the sensitive GRS unit. They show preliminary experimental results using a grating beam splitter demonstrating an optical readout sensitivity of 30 pm $Hz^{-1/2}$. In [17] in order to minimize low-frequency instabilities they choose a homodyne technique, to avoid moving or active elements. As a drag-free sensor one of the problems of the applicability of interferometry is that displacements at a given time are measured with respect to some initial position. In that case power losses or rapid motion of the spacecraft could produce error in counting fringes and therefore in the distance measurements. To solve this problem they propose to use a wavelength modulation technique. A prototype bench top interferometer based on this technique is shown to achieve a shot-limited displacement sensitivity of $3 \times 10^{-12}$ m $Hz^{-1/2}$ above 60 Hz. Finally in [18] a concept of laser metrology inertial sensor was suggested in which transmit/receive laser lights are reflected by the proof mass, the metrology is based on sources with different wavelengths and light pressure is used for active control of the proof mass.

# 5 Proof mass-spacecraft（PM-SC）stiffness coupling

Magnetic and electrostatic interactions (spacecraft magnetic gradient, patch fields, proof mass charging, applied voltages for capacitive sensing, etc) and local gravity gradients are responsible for coupling the proof mass to the spacecraft motion. Different contributions to the proof mass-spacecraft stiffness are listed in table 6. The first four contributions have to be considered in the case of employing capacitive sensing. Stiffness due to patch potentials is the dominant contribution with a value of the order of $10^{-9}$ $s^{-2}$, where a worst case value for the standard deviation of the patch potential distribution of $V_{pe} \approx 0.1$ V has been assumed. All these contributions can be suppressed by increasing the gap between the proof mass and the surrounding electrodes, *d*.

In the case of optical sensing there is no need for using close gaps between the proof mass and the entire spacecraft structure. Therefore the contribution due to patch fields would be negligible. Moreover optical sensing can be in principle considered stiffness free.

Stiffness due to local gravity gradients is common to both ways of sensing, capacitive and optical. In [19] the stiffness due to local gravity gradients is worked out considering a cylindrical spacecraft structure and a rectangular proof mass, giving a spacecraft-proof mass stiffness about $3.5 \times 10^{-10}$ $s^{-2}$. In the case of using capacitive sensing the sensing electrodes will also be a source of local gravity gradients. Again in [19] it has been estimated a value of $5.7 \times 10^{-8}$ $s^{-2}$ for electrodes-proof mass stiffness, where it is also mentioned that the local gravity gradients of the surrounding electrodes could be minimized by using a cubical proof mass and capacitive enclosure.

In summary, patch fields, in the case of capacitive sensing, and local gravity gradients, in both optical and capacitive sensing, are the leading contributions to the stiffness. Taking into account the previous discussion it can be considered a preliminary stiffness value about $3.5 \times 10^{-10}$ $s^{-2}$ for optical sensing, and $5.7 \times 10^{-8}$ $s^{-2}$ for capacitive sensing, keeping in mind that the correct values for ASTROD need to be worked out by appropriate modeling of the spacecraft and payload.

## 6 ASTROD noise requirements at 100 μHz

Aiming for a ten-fold improvement in acceleration noise sensitivity with respect to LISA, ASTROD would be able to test relativistic gravity to 1 ppb. The LISA observational bandwidth is $10^{-4}$ Hz $\leq f \leq 1$ Hz with an acceleration noise sensitivity target of $3 \times 10^{-15} [1 + (f/3\ \text{mHz})^2]$ m s$^{-2}$ Hz$^{-1/2}$. Possibilities to extend the LISA observational bandwidth to lower frequencies and its impact in the study of some potential astrophysical sources has been discussed in the literature [13, 20]. Acceleration noise targets of LISA at low frequencies of $3 \times 10^{-15} (10^{-4}\text{Hz}/f)^{1/2}$ m s$^{-2}$ Hz$^{-1/2}$ in the frequency range $10^{-5}$ Hz $\leq f \leq 10^{-4}$ Hz and $9.5 \times 10^{-15} (10^{-5}\text{Hz}/f)$ m s$^{-2}$ Hz$^{-1/2}$ in the frequency range $3 \times 10^{-6}$ Hz $\leq f \leq 10^{-5}$ Hz have been suggested in [13].

Inspection of table 2 shows that reducing the residual pressure by a factor 3, from $3 \times 10^{-6}$ to $10^{-6}$ Pa, using a factor 10 in magnetic shielding, $\xi_m$, a factor 100 in electrostatic shielding, $\xi_e$, a factor 150 in the optical bench thermal shielding factor, $\xi_{TS}$, and also assuming a discharged proof mass we have that the environmental direct proof mass acceleration disturbances are of the order of $1.7 \times 10^{-16}$ m s$^{-2}$ Hz$^{-1/2}$. Back action disturbances associated to capacitive sensing are listed in table 4. Again by continuously discharging the proof mass, disturbances proportional to the total charge, $f_{\delta q,2}$ and $f_{\delta V,2}$, would be suppressed to a great extend. On the other hand, dielectric losses, $f_{DL}$, and disturbance associated to charge fluctuations, $f_{\delta q,1}$ are the most relevant contributions. Both can be reduced by increasing the capacitance gaps. In the case of dielectric losses, given a gap of 4 mm, a loss angle, $\delta \sim 10^{-6}$ and a dc bias voltage, $V_0 \approx 10^{-2}$ V, we obtain an acceleration disturbance of $1.8 \times 10^{-17}$ m s$^{-2}$ Hz$^{-1/2}$. Keeping the same gap of 4 mm, and assuming the same effective charging and discharging rate we need a voltage difference across opposite electrodes, $V_d \leq 1$ mV, to be just under the target sensitivity of $3 \times 10^{-16}$ m s$^{-2}$ Hz$^{-1/2}$. We have assumed a voltage difference of 0.5 mV. The total back action acceleration disturbance is then of the order of $1.4 \times 10^{-16}$ m s$^{-2}$ Hz$^{-1/2}$, giving a total direct acceleration disturbance when using capacitive sensing of $f_p \approx 2.2 \times 10^{-16}$ m s$^{-2}$ Hz$^{-1/2}$ at 0.1 mHz.

One of the advantages of optical sensing readout is that back action acceleration disturbances can be made negligible. In that case, total direct acceleration disturbances would be of the order of $1.7 \times 10^{-16}$ m s$^{-2}$ Hz$^{-1/2}$ at 0.1 mHz. In the case of sub-picometer shot noise limited sensitivity with a counter balance force to 0.1%, the back action force on to a 1.75 kg mass is of the order of $4 \times 10^{-18}$ m s$^{-2}$ Hz$^{-1/2}$, well below our target sensitivity. Acceleration disturbance due to readout noise coupled via stiffness could in principle be considered negligible in the case of optical sensing. In both cases, when optical or capacitive sensing is employed, local gravity gradients are a major source of stiffness. In the case of optical sensing we attempt to use the value $3.5 \times 10^{-10}$ s$^{-2}$ reported before. Considering optical readout sensitivity of the order of $10^{-10}$ m Hz$^{-1/2}$ it shows that for optical sensing this contribution would be negligible. This contribution is more important when employing capacitive sensing. Assuming a stiffness value of $5.7 \times 10^{-8}$ s$^{-2}$ ($4 \times 10^{-7}$ s$^{-2}$ requirement for LISA) we would require a readout sensitivity about $2 \times 10^{-9}$ m Hz$^{-1/2}$ to obtain similar noise levels than direct acceleration disturbances. External disturbances also contribute via stiffness coupling. In the case of capacitive and optical sensing this contribution is $f_p^{cap} \approx 5.1 \times 10^{-10}/u$ and $f_p^{opt} \approx 3.1 \times 10^{-11}/u$, respectively. If we aim for an acceleration noise of the order of $10^{-16}$ m s$^{-2}$ Hz$^{-1/2}$ we need to achieve control loop gains of the order of $u \geq 5 \times 10^6$ and $u \geq 3 \times 10^5$ for the case of capacitive and optical sensing respectively.

# 7 ASTROD below 100 μHz. Gravitational wave responses

The significance of low-frequency gravitational-wave response in the study of certain astrophysical sources, like massive black hole (MBH) binaries at high red shift, has been discussed in the literature [13, 20]. For instance, the possibility of extending the LISA observational bandwidth to lower frequencies has been proposed in [13], suggesting acceleration noise targets of $3 \times 10^{-15} (10^{-4}\,\text{Hz}/f)^{1/2}$ in the frequency range $10^{-5}$ Hz $\leq f \leq 10^{-4}$ Hz and $9.5 \times 10^{-15} (10^{-5}\,\text{Hz}/f)$ in the frequency range $3\times10^{-6}$ Hz $\leq f \leq 10^{-5}$ Hz.

Table 7 shows the contribution of non-thermal, thermal and sensor back action disturbances to the acceleration noise at frequencies $3 \times 10^{-5}$, $10^{-5}$ and $3 \times 10^{-6}$ Hz, considering the parameters given in table 5. Magnetic interaction of the solar field with the proof mass magnetic susceptibility, $f_{m2}$, is the dominant contribution of the non-thermal disturbances. That is also the case of thermal radiation pressure, $f_{TR}$, within the thermal disturbances. A significant improvement in active/passive thermal isolation at those frequencies would suppress thermal radiation pressure, leaving thermally induced gravity gradients as the ultimate limiting source of thermal disturbance. Nevertheless to evaluate this disturbance would require accurate gravitational and thermal modeling.

At low frequency certain sensor back action disturbances are of importance. Disturbances due to dielectric losses, $f_{DL}$, and voltage difference between opposite electrodes, $f_{\delta q,1}$, are significant noise sources when employing capacitive sensing. It is worth noticing that unless a significant improvement in passive/active thermal isolation is achieved, in order to suppress thermal radiation pressure, the choice of capacitive or optical sensing will not be so relevant at such frequencies. Figure 1 shows the acceleration noise sensitivity requirement for ASTROD compared to LISA, and its extension to low frequencies proposed by Bender [13]. By inspection of table 7, to achieve ASTROD accelerometer noise requirement below $10^{-5}$ Hz, particularly at $3 \times 10^{-6}$ Hz, we would need improvements on active thermal isolation and, in the case of employing capacitive sensing, we would need a significant reduction in the stray DC voltage difference across opposite electrodes of the sensor.

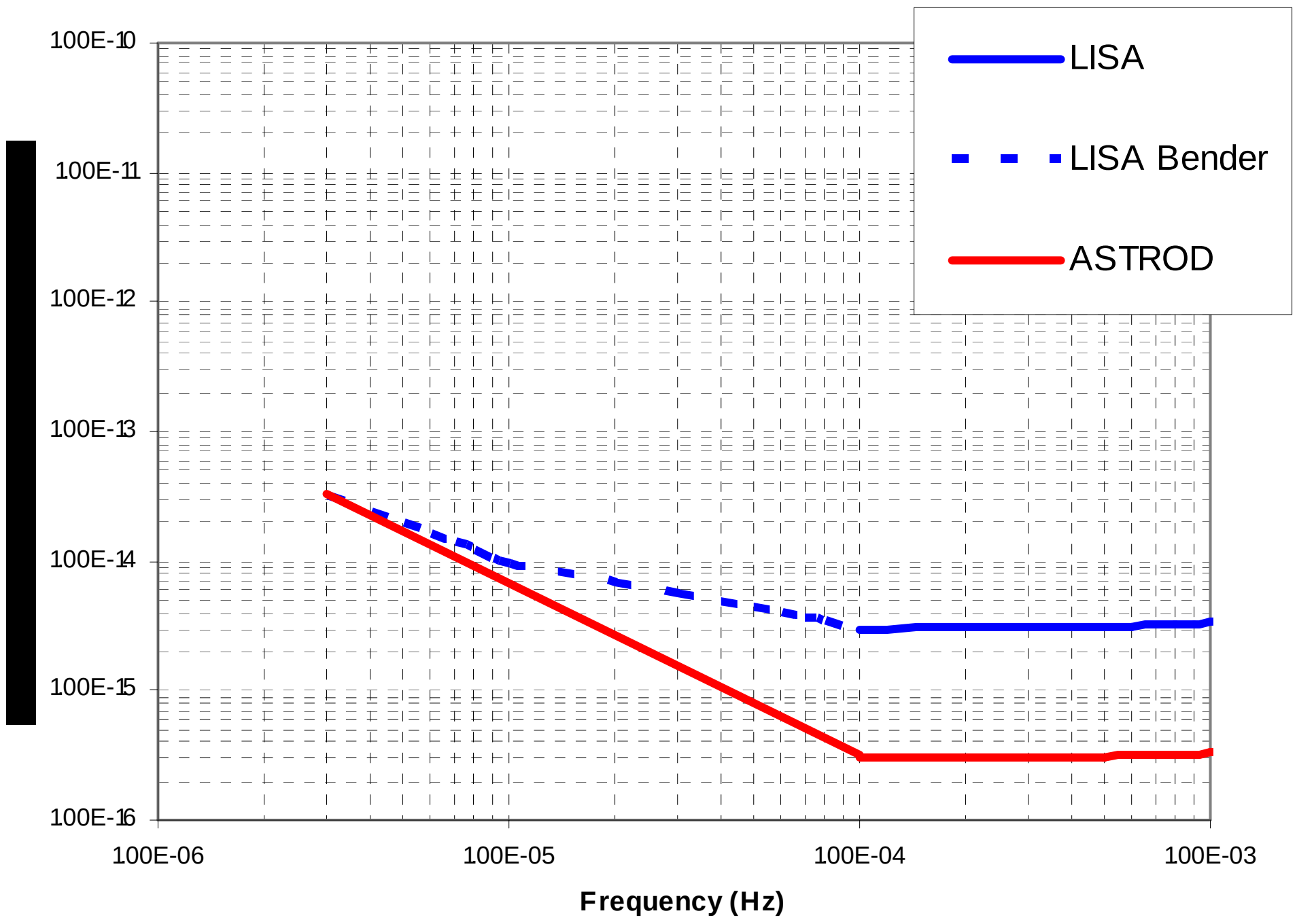

Figure 1. Acceleration noise sensitivity for LISA and ASTROD.

To obtain the gravitational wave strain sensitivity we have to consider acceleration disturbances and also optical path errors. Optical path noise is caused by frequency instability of lasers, shot noise, etc. Shot noise is inversely proportional to $P_r^{1/2}$, where $P_r$ is the received power. It is also proportional to the distance between the proof masses. Therefore strain sensitivity due to shot noise is independent of the interferometer arm length giving us the baseline of the gravitational wave sensitivity curve. On the other hand the strain sensitivity due to acceleration disturbances improves in proportion to the arm length. ASTROD mission proposes a gravitational wave interferometer of varying arm length. Because of longer arm length, in average 30 times longer than those of LISA, the sensitivity curve for ASTROD is shifted to lower frequencies.

Figure 2 shows the gravitational wave sensitivity for ASTROD and LISA compared to gravitational strain of different black hole binary merges. To estimate the gravitational wave strength sensitivity we make use of the expression for a Michelson-type interferometer with equal arms, which is given by,

$$\sqrt{S_h^{M0}(f)} \approx \frac{1}{\sin cu_0} rss\left[\frac{4}{\pi^2}\left(\frac{hc}{\eta P_t}\right)^{1/2}\frac{\lambda^{3/2}}{D^2}, \frac{2A_0}{L(2\pi f)^2}\right] \mathrm{Hz}^{-1/2} \qquad (3.14)$$

where $u_0 \equiv \omega L/c$ and $A_0$ is the proof mass acceleration noise. For ASTROD, assuming a laser power of 10 W, 1 μm lasers, 30% overall optical efficiency, η, and 30 centimeters aperture, D, the shot noise level is approximately $1.2 \times 10^{-21}$ $\mathrm{Hz}^{-1/2}$.

To calculate gravitational wave strain for massive black hole binaries we follow [20]. The characteristic strain of black hole binary merges can be expressed by,

$$h_c = 6.5\times 10^{-17}\left(\frac{D_L(z)}{1Gpc}\right)^{-1}\left(\frac{(1+z)(M^2\mu^3)^{1/5}}{10^6 M_S}\right)^{5/6}\left(\frac{f}{10^{-4}Hz}\right)^{-1/6} \qquad (3.15)$$

where $D_L(z)$ is the distance luminosity, z is the red shift, $\mu$ and $M_S$ are the reduced and total mass of the binary. The luminosity distance is given by,

$$D_L(z) = 3(1+z)\frac{1}{h}\int_0^z \frac{dz'}{\left[\Omega_M(1+z')^3 + (1-\Omega_M)\right]^{1/2}} \qquad (3.16)$$

in Gpc, where values of $\Omega_M$=0.27 and h=0.71 have been considered.

We also have to take into account that the frequency of the binary is not the same that the frequency observed by the detector. Following [20] the observed frequency by the detector is given by,

$$f = 5.4 \times 10^{-5}\, Hz \left( \frac{\left(M^2 \mu^3\right)^{1/5} (1+z)}{10^6 M_S} \right)^{-5/8} \left( \frac{t_{obs}}{6\ months} \right)^{-3/8} \qquad (3.17)$$

Gravitational wave strains for different black hole binary merges are shown in figure 2 compared with the gravitational strain sensitivity for ASTROD and LISA.

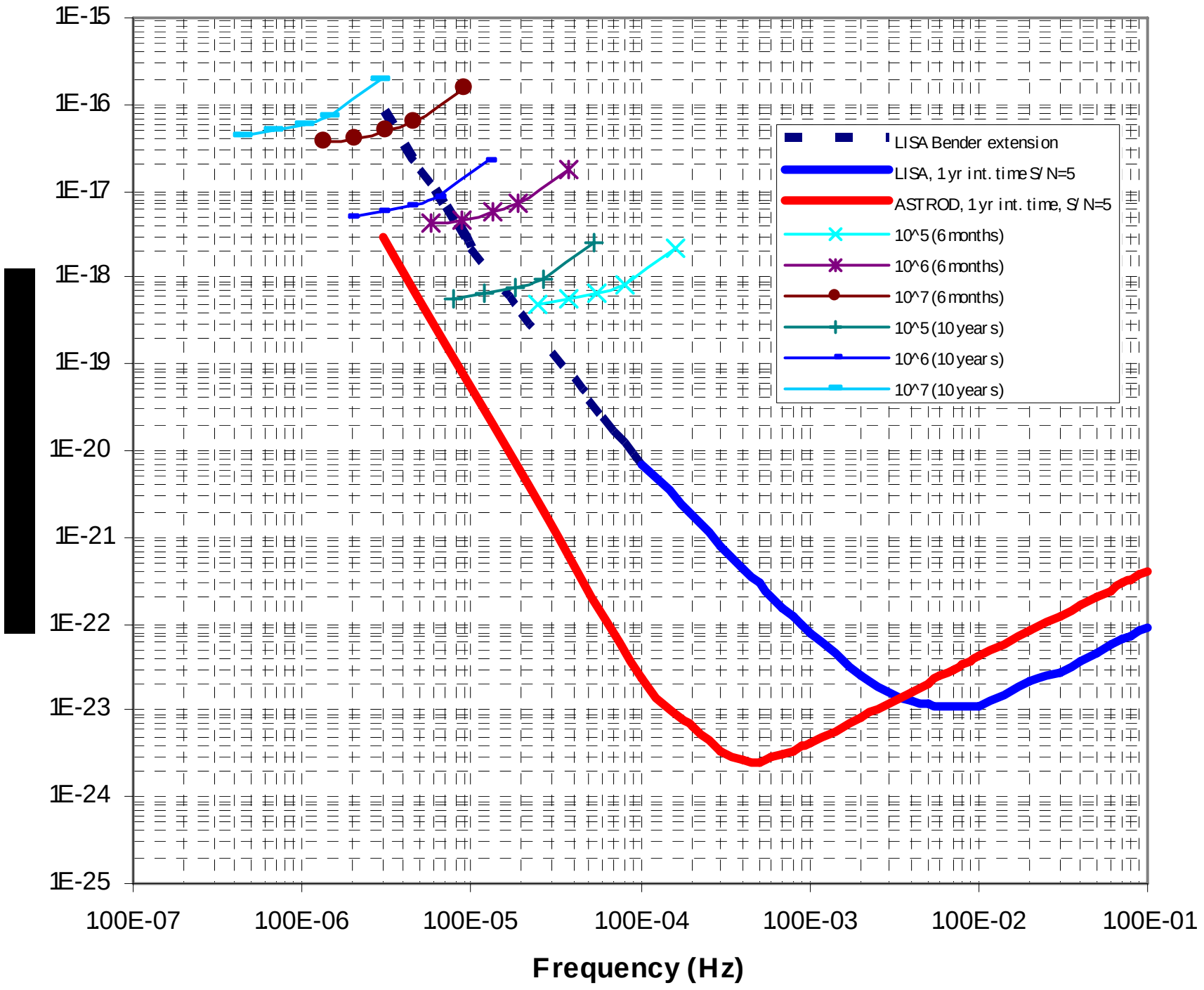

Figure 2. Gravitational wave strengths of massive black hole binaries at different red shifts (z=1, 5, 10, 20 and 40), 6 months and 10 years before coalescence, compared to ASTROD and LISA GW strain sensitivities. It has been considered 1 year integration time and S/N of 5.

# 8 Conclusions

LISA acceleration noise target sensitivity is $3 \times 10^{-15}$ m $s^{-2}$ $Hz^{-1/2}$ at 100 μHz. ASTROD aims to improve LISA accelerometer noise by a factor 3-10. A ten-fold improvement would allow us to test relativistic gravity to 1 ppb. We have tentatively discussed ways of improving acceleration noise by reducing housing pressure, improving magnetic, electrostatic and thermal shielding, discharging schemes, etc. We have also compared the performance of optical and capacitive sensing. Optical sensing is, in principle, a stiffness-free sensor exerting very low back action force. Finally we have obtained a gravitational wave response curve for the given ASTROD acceleration noise and we discussed its impact in the study of certain astrophysical sources like MBH binaries at high red shift.

ACKNOWLEDGEMENTS

We thank the National Natural Science Foundation (Grant No 10475114) and the Foundation of Minor Planets for funding this work.

# Table 1

| Spacecraft acc. Disturbances | Expressions | 0.1 mHz |
|---|---|---|
| Solar radiation pressure | $f_{ns} = \frac{2A_{SC}\delta W_0}{M_{SC}c}$ | $10^{-9}$ |
| Thruster noise | $1\,\mu\mathrm{NHz}^{-1/2}$ | $2\times10^{-9}$ |

Table 1. Spacecraft acceleration disturbances at 0.1 mHz.

# Table 2

| Sources of disturbances | Expressions | Frequency dependence | Noise (units $10^{-16}$ m s$^{-2}$ Hz$^{-1/2}$) |
|---|---|---|---|
| Cosmic rays | $f_{CR} = \frac{\sqrt{2mE\lambda}}{m_P}$ | $1.5\times10^{-2}$ | $1.5\times10^{-2}$ |
| Residual gas | $f_{RG} = \frac{\sqrt{2PA_P}}{m_P}\left(3k_BT_Pm_N\right)^{1/4}$ | $2.8\left(\frac{P}{3\times10^{-6}}\right)^{1/2}$ | 1.6 |
| *Magnetic susceptibility I (χ).* | $f_{m1} = \frac{2\chi}{\mu_0\rho}\frac{1}{\xi_m}\delta B_{SC}\nabla B_{SC}$ | $0.72\left(\frac{\chi}{3\times10^{-6}}\right)\frac{1}{\xi_m}$ | 0.072 |
| Magnetic susceptibility II (χ) | $f_{m2} = \frac{\sqrt{2}\chi}{\mu_0\rho}\frac{1}{\xi_m}\nabla B_{SC}\delta B_{IP}$ | $2.0\frac{1}{\xi_m}\left(\frac{\chi}{3\times10^{-6}}\right)\left(\frac{0.1\,\mathrm{mHz}}{f}\right)^{2/3}$ | 0.20 |
| Permanent magnetic moment | $f_{m3} = \frac{1}{\sqrt{2m_P\xi_m}}\lvert M_r\rvert\lvert\nabla(\delta B)\rvert$ | $3.2\frac{1}{\xi_m}\left(\frac{\lvert M_r\rvert}{2\times10^{-8}}\right)$ | 0.32 |
| Lorentz I. | $f_{L1} = \frac{v}{m_P}\frac{1}{\xi_e}q\delta B_{IP}$ | $9.1\times10^{-2}\left(\frac{\bar{\dot{Q}}t}{10^{-13}}\right)\left(\frac{100}{\xi_e}\right)\left(\frac{0.1\,\mathrm{mHz}}{f}\right)^{2/3}$ | $9.1\times10^{-3}$ |
| Lorentz II. | $f_{L2} = \frac{v}{m_P}\frac{1}{\xi_e}B_{IP}\delta q$ | $1.7\times10^{-3}\left(\frac{\bar{\dot{Q}}}{288}\right)^{1/2}\left(\frac{100}{\xi_e}\right)\left(\frac{0.1\,\mathrm{mHz}}{f}\right)$ | $2\times1.7\times10^{-3}$ |
| Radiometer effec**t** | $f_{RE} = \frac{A_PP}{2m_P}\frac{1}{\xi_{TS}}\frac{\delta T_{OB}}{T_P}$ | $4.7\left(\frac{P}{3\times10^{-6}}\right)\frac{1}{\xi_{TS}}$ | $1.0\times10^{-2}$ |
| Out gassing effect | $f_{OG} = 10\,f_{RE}$ | $47\left(\frac{P}{3\times10^{-6}}\right)\frac{1}{\xi_{TS}}$ | $1.0\times10^{-1}$ |
| Thermal radiation pressure | $f_{TR} = \frac{8\sigma}{m_P}\frac{A_P}{c}T_P^3\frac{\delta T_{OB}}{\xi_{TS}}$ | $12\frac{1}{\xi_{TS}}$ | 0.08 |
| Gravity Gradient | $f_{GG} = \frac{2GM}{r^2}\alpha\delta T_{SC}$ | $0.54\left(\frac{M}{1\,\mathrm{kg}}\right)\left(\frac{\delta T_{SC}}{0.004\,\mathrm{KHz}^{-1/2}}\right)$ | 0.54 |
| **Total proof mass acceleration noise at 0.1 mHz (m s$^{-2}$ Hz$^{-1/2}$)** | | | 1.7 |

Table 2. Proof mass direct environmental disturbances at 0.1 mHz

## Table 3

| Frequency (Hz) | $\delta T_{OB}$ (K Hz$^{-1/2}$) | $f_{RE}$ (m s$^{-2}$ Hz$^{-1/2}$) | $f_{TR}$ (m s$^{-2}$ Hz$^{-1/2}$) |
|---|---|---|---|
| $f \geq 10^{-4}$ | $3.0\times10^{-5}\left(\frac{1\,\text{mHz}}{f}\right)^{1/2}$ | $4.7\times10^{-16}\frac{1}{\xi_{TS}}\left(\frac{P}{3\times10^{-6}}\right)\left(\frac{0.1\,\text{mHz}}{f}\right)^{1/2}$ | $1.2\times10^{-15}\frac{1}{\xi_{TS}}\left(\frac{0.1\,\text{mHz}}{f}\right)^{1/2}$ |
| $3\times10^{-5}$ | $32\times10^{-5}$ | $1.6\times10^{-15}\frac{1}{\xi_{TS}}\left(\frac{P}{3\times10^{-6}}\right)$ | $1.2\times10^{-14}\frac{1}{\xi_{TS}}$ |
| $10^{-5}$ | $12\times10^{-3}$ | $6.1\times10^{-14}\frac{1}{\xi_{TS}}\left(\frac{P}{3\times10^{-6}}\right)$ | $4.6\times10^{-13}\frac{1}{\xi_{TS}}$ |
| $3\times10^{-6}$ | $13\times10^{-2}$ | $6.6\times10^{-13}\frac{1}{\xi_{TS}}\left(\frac{P}{3\times10^{-6}}\right)$ | $4.9\times10^{-12}\frac{1}{\xi_{TS}}$ |

Table 3. Thermal disturbances at low frequencies, below 0.1 mHz.

## Table 4

| Source of disturbances | Expressions | Frequency dependence | Noise in units ($10^{-16}$ m s$^{-2}$ Hz$^{-1/2}$) |
|---|---|---|---|
| Quantization | $f_q = \frac{10\lvert F_{x0}\rvert}{m_P}\frac{1}{2^N}\frac{1}{\sqrt{12\nu_s}}$ | $6.3\times10^{-4}\left(\frac{V_d}{5\times10^{-3}}\right)\left(\frac{V_{x0}}{10^{-2}}\right)$ | $6.3\times10^{-4}$ |
| Dielectric losses | $f_{DL} = \frac{\sqrt{2}C_x}{m_P d}V_0\delta v_{diel}$ | $2.5\left(\frac{\delta}{10^{-5}}\right)^{1/2}\left(\frac{V_0}{10^{-2}}\right)\left(\frac{4\times10^{-3}}{d}\right)^{3/2}\left(\frac{0.1\,\text{mHz}}{f}\right)^{1/2}$ | 0.18 |
| Voltage | $f_{\delta V_d,1} = \frac{C_x}{m_P d}\frac{C_x}{C}\left(V_{x0}-V_g\right)\delta V_d$ | $0.145\left(\frac{V_{0g}}{10^{-2}}\right)\left(\frac{\delta V_d}{10^{-5}}\right)\left(\frac{4\times10^{-3}}{d}\right)^2$ | 0.07 |
| Charging-Voltage 1 | $f_{\delta V_d,2} = \frac{q}{dm_P}\frac{C_x}{C}\delta V_d$ | $0.24\left(\frac{4\times10^{-3}}{d}\right)\left(\frac{q}{10^{-13}}\right)\left(\frac{\delta V_d}{10^{-5}}\right)$ | 0.024 |
| Charging-Voltage 2 | $f_{\delta q,1} = \frac{1}{dm_P}\frac{C_x}{C}V_d\delta q$ | $2\times7\left(\frac{4\times10^{-3}}{d}\right)\left(\frac{V_d}{5\times10^{-3}}\right)\left(\frac{\bar{\dot{Q}}}{288}\right)^{1/2}\left(\frac{0.1\,\text{mHz}}{f}\right)$ | 1.4 |
| Charging | $f_{\delta q,2} = \frac{q}{m_P d^2}\frac{C_x}{C^2}\Delta d\delta q$ | $0.01\left(\frac{q}{10^{-13}}\right)\left(\frac{\Delta d}{10\,\mu\text{m}}\right)\left(\frac{\bar{\dot{Q}}}{288}\right)^{1/2}\left(\frac{4\times10^{-3}}{d}\right)\left(\frac{0.1\,\text{mHz}}{f}\right)$ | $4\times10^{-3}$ |
| **Total sensor back action disturbance (capacitance)** | | | 1.4 |

Table 4. Capacitive sensing back action disturbances at 0.1 mHz.

## Table 5

| **Paraneter values used in the acceleration noise estimates.** | |
|---|---|
| **Proof Mass** | |
| Mass (kg) | 1.75 |

| | | |
|---|---|---|
| Density ($kgm^{-3}$) | | $2\times10^4$ |
| Cross Section ($m^2$) | | $0.050\times0.035$ |
| Temperature (K) | | 293 |
| Magnetic Susceptibility: $\chi$ | | $3\times10^{-6}$ |
| Permanent Magnetic Moment: $M_r$ ($Am^2Kg^{-1}$) | | $2\times10^{-8}$ |
| Maximum charge build-up: | | UV light continuous discharging |
| Velocity [$ms^{-1}$] | | $4\times10^4$ |
| Electrostatic shielding factor $\xi_e$ | | 100 |
| Magnetic shielding factor $\xi_m$ | | 10 |
| Optical bench thermal shielding factor $\xi_{TS}$ | | 150 |
| Residual gas pressure | | $10^{-6}$ |
| **Magnetic fields.** | | |
| Local Magnetic field | $B_{SC}$ $[T]$ | $8\times10^{-7}$ |
| Local Magnetic field gradient | $\nabla B_{SC}$ $[Tm^{-1}]$ | $3\times10^{-6}$ |
| Fluctuation in local magnetic field | $\delta B_{SC}$ $[THz^{-1/2}]$ | $1\times10^{-7}$ |
| Interplanetary magnetic field | $B_{ip}$ $[T]$ | $1.2\times10^{-7}$ |
| Interplanetary magnetic field gradient | $\delta B_{ip}$ $[THz^{-1/2}]$ | $4\times10^{-7}\left(0.1\text{mHz}/f\right)^{2/3}$ |
| Gradient of time-varying magnetic field | $\nabla(\delta B)$ $[Tm^{-1}Hz^{-1/2}]$ | $4\times10^{-8}$ |
| **Capacitive sensing** | | |
| Capacitance $C_x$ [pF] | | 3 |
| Capacitance to ground $C_g$ [pF] | | 3 |
| Total capacitance C [pF]($\approx$ $6C_x$) | | 18 |
| Gap d [mm] | | 4 |
| Proof mass bias voltage $V_{M0}$ [V] | | 0.6 |
| Voltage difference to ground $V_{x0}$-$V_g$=$V_{0g}$ [V] | | 0.01 |
| Voltage difference between opposite faces $V_d$ [V] | | $5\times10^{-4}$ |
| Fluctuation voltage difference $\delta V_d$ [V $Hz^{-1/2}$] | | $10^{-5}$ |
| Residual dc bias voltage on electrodes $V_0$ [V] | | $10^{-2}$ |
| Loss angle $\delta$ | | $10^{-6}$ |
| Gap asymmetry $\Delta$d [$\mu$m] | | 10 |
| **Quantization** | | |
| Net force on the proof mass: $F_{x0}$ [N] | | $2.5\times10^{-14}$ |
| Binary digit: N [bits] | | 16 |
| Sampling frequency: $\nu_s$ [Hz} | | 100 |

Table 5. Parameter values for ASTROD acceleration noise estimates.

# Table 6

| PM-spacecraft stiffness | Expressions | ASTROD ($s^{-2}$) | 0.1 mHz |
|---|---|---|---|
| Image charges | $K_c=\frac{q^2}{d^2m_p}\frac{C_x}{C^2}$ | $3.3\times10^{-12}\left(\frac{q}{10^{-13}}\right)^2\left(\frac{4\times10^{-3}}{d}\right)$ | $3.3\times10^{-14}$ |
| Applied voltage | $K_V = \frac{C_x}{m_p d^2}\left[\left(\frac{C_x}{C}+\frac{1}{4}\right)V_d^2 + \left(\frac{C_g}{C}\right)^2 V_{0g}^2\right]$ | $1.1\times10^{-13}\left(\frac{4\times10^{-3}}{d}\right)^3\left[10.4\left(\frac{V_d}{5\times10^{-3}}\right)^2 + 2.8\left(\frac{V_{0g}}{10^{-2}}\right)^2\right]$ | $3.2\times10^{-13}$ |
| $q\times V_{0g}$ | $K_{CV} = \left(\frac{2}{m_p d^2}\right)\left(\frac{C_x}{C}\right)\left(\frac{C_g}{C}\right)qV_{0g}$ | $4\times10^{-12}\left(\frac{4\times10^{-3}}{d}\right)\left(\frac{q}{10^{-13}}\right)\left(\frac{V_{0g}}{10^{-2}}\right)$ | $4\times10^{-13}$ |
| Patch fields | $K_{PF} = \gamma\frac{C_x}{m_p d^2}\left(\frac{C_x}{C}\right)^2 V_{pe}^2$ | $1.5\times10^{-10}\left(\frac{4\times10^{-3}}{d}\right)^3\left(\frac{V_{pe}}{0.1}\right)$ | $1.5\times10^{-10}$ |
| Gravity Gradient | $K_{GG} = \frac{2GM}{r^3}$ | $3.2\times10^{-10}\left(\frac{M_{dis}}{1kg}\right)\left(\frac{0.75}{r}\right)^3$ | $3.2\times10^{-10}$ |
| Induced magnetic moment | $K_{m1} = \frac{2\chi}{\rho\mu_0}\left[\left\|\nabla B_{SC}\right\|^2 + B_{SC}\left\|\nabla^2 B_{SC}\right\|\right]$ | $5.4\times10^{-15}\left(\frac{\chi}{3\times10^{-6}}\right)$ | $5.4\times10^{-15}$ |
| Magnetic remanent moment | $K_{m2} = \frac{1}{\sqrt{2}m_P}\left\|M_r\right\|\left\|\nabla^2 B_{SC}\right\|$ | $7.9\times10^{-14}\left(\frac{\left\|M_r\right\|}{1.1\times10^{-8}}\right)$ | $7.9\times10^{-14}$ |
| Total stiffness (Capacitance) | | | $3.5\times10^{-10}$ |

Table 6. Proof mass-spacecraft stiffness terms. Capacitive sensing is considered.

# Table 7

| **Frequency (Hz)** | **Bender LISA ($m\ s^{-2}\ Hz^{-1/2}$)** | **Thermal disturbances ($m\ s^{-2}\ Hz^{-1/2}$)** | **Non- thermal disturbances ($m\ s^{-2}\ Hz^{-1/2}$)** | **Sensor back action (Capacitive sensor) ($m\ s^{-2}\ Hz^{-1/2}$)** |
|---|---|---|---|---|
| $3\times10^{-5}$ | $5.5\times10^{-15}$ | $8.8\times10^{-17}(150/\xi_{TS})$ | $4.5\times10^{-17}$ | $4.7\times10^{-16}$ |
| $10^{-5}$ | $9.5\times10^{-15}$ | $3.3\times10^{-15}(150/\xi_{TS})$ | $9.3\times10^{-17}$ | $1.4\times10^{-15}$ |
| $3\times10^{-6}$ | $3.2\times10^{-14}$ | $3.6\times10^{-14}(150/\xi_{TS})$ | $2.0\times10^{-16}$ | $4.7\times10^{-14}$ |

Table 7. Contribution to the proof mass acceleration noise at low frequency due to thermal, non-thermal, and sensor back action frequency dependent disturbances.